\documentclass[pre,onecolumn,superscriptaddress,nofootinbib]{revtex4}

\usepackage{graphicx}
\usepackage{dcolumn}
\usepackage{array}
\usepackage{tabularx}
\usepackage{color}

\usepackage{floatrow}

\usepackage{eucal}
\usepackage[dvips]{epsfig}
\usepackage{amssymb}
\usepackage{amsfonts}

\usepackage{amsmath,amsthm}

\newcommand{\be}{\begin{eqnarray}}
\newcommand{\ee}{\end{eqnarray}}
\newcommand{\bse}{\begin{subequations}}
\newcommand{\ese}{\end{subequations}}

\newcommand{\bnum}{\begin{enumerate}}
\newcommand{\enum}{\end{enumerate}}

\newcommand{\bit}{\begin{itemize}}
\newcommand{\eit}{\end{itemize}}

\newcommand{\bc}{\begin{cases}}
\newcommand{\ec}{\end{cases}}

\newcommand{\bpm}{\begin{pmatrix}}
\newcommand{\epm}{\end{pmatrix}}

\newcommand{\bvm}{\begin{vmatrix}}
\newcommand{\evm}{\end{vmatrix}}

\newcommand{\bs}{\boldsymbol}

\newcommand{\eps}{\epsilon}

\newcommand{\gs}{\sigma}

\newcommand{\Go}{\Omega}

\newcommand{\Gl}{\Lambda}

\newcommand{\p}{\partial}
\newcommand{\f}{\frac}

\newcommand{\tn}{\textnormal}

\begin{document}

\title{Stokes' second problem and reduction of inertia in active fluids}

\author{Jonasz S\l{}omka}
\affiliation{Department of Mathematics, Massachusetts Institute of Technology, 77 Massachusetts Avenue, Cambridge, MA 02139-4307, USA}

\author{Alex Townsend}
\affiliation{Department of Mathematics, Cornell University, Malott Hall, Ithaca, NY 14853, USA}

\author{J{\"o}rn Dunkel}
\affiliation{Department of Mathematics, Massachusetts Institute of Technology, 77 Massachusetts Avenue, Cambridge, MA 02139-4307, USA}

\date{\today}
\begin{abstract}
We study a generalized Navier-Stokes model describing the thin-film flows in non-dilute suspensions of ATP-driven microtubules or swimming bacteria that are enclosed by a moving ring-shaped container. Considering Stokes' second problem, which concerns the motion of an oscillating boundary,  our numerical analysis predicts that a periodically rotating ring will oscillate at a higher frequency in an active fluid than in a passive fluid, due to an  activity-induced reduction of the fluid inertia.  In the case of a freely suspended fluid-container system that is isolated from external forces or torques, active fluid stresses can induce large fluctuations in the container's angular momentum if the confinement radius matches certain multiples of the intrinsic vortex size of the active suspension. This effect could be utilized to transform collective microscopic swimmer activity into macroscopic motion in optimally tuned geometries.

\end{abstract}


\maketitle


\section{Introduction}

Pendulums swinging in air or water exhibit periods longer than those predicted based on gravity and buoyancy~\cite{stokes1851effect,nelson1986pendulum}. In his famous mid-19th century work~\cite{stokes1851effect}, George G. Stokes resolved the discrepancy by postulating an additional parameter, the \lq index of friction\rq{}  (viscosity), in the hydrodynamic equations that now bear his name. Building on this insight, Stokes was able to calculate the terminal velocity of sedimenting globules set by the viscous drag, providing a partial explanation for the suspension of clouds~\cite{hinds2012aerosol}. Since then, the term  \textit{Stokes' problems} (SPs) has become synonymous with the investigation of objects that move either uniformly or in an oscillatory manner through a liquid~\cite{wiggins1998trapping,zeng1995stokes}. Nowadays, the traditional SPs provide important reference points for the  rheology  of active fluids, such as water-based solutions driven by swimming bacteria~\cite{2007SoEtAl,Dunkel2013_PRL} or microtubule networks~\cite{2012Sanchez_Nature,2015Giomi}. Recent experiments show that sufficiently dense bacterial suspensions can significantly reduce the drag experienced by a moving sphere~\cite{2009SokolovAranson} or rotated cylindrical walls~\cite{2015LoGaDoAuCle}. Several theories have been proposed to rationalize the observed decrease in shear viscosity,  ranging from microscopic and Fokker--Planck-based approaches~\cite{haines2009three,saintillan2010dilute,ryan2011viscosity,PhysRevLett.118.018003} for dilute suspensions to active liquid crystal continuum models~\cite{Hatwalne2004prl,2008CaFieMarOrYoe,giomi2010sheared,PhysRevE.83.041910,2012Foffano_EPJE} and phenomenologically generalized Navier--Stokes equations  for non-dilute suspensions~\cite{2017SlomkaDunkel_PRF}. By contrast, the effects of oscillatory boundary conditions -- Stokes' second problem -- have thus far only been partially explored in dilute active fluids~\cite{Guo2017_SoR}. Therefore, it is currently unknown how the collective microbial swimming dynamics in dense suspensions, which typically exhibit active turbulence with characteristic vortex length scale $\Gl$ and correlation time~$\tau$~\cite{2004DoEtAl,2007Cisneros,2012Sanchez_Nature,2012Sokolov,Dunkel2013_PRL,sumino2012nature,Wioland2013_PRL}, interacts with oscillating boundaries. In particular, it is not known how the frequency of a pendulum  is altered by the presence of an active fluid component. Here, we will show that activity effectively reduces the fluid inertia, thus increasing the frequency relative to that of an identical pendulum swinging in water.

\begin{figure}[t]
\centering
\includegraphics[width=0.35\textwidth]{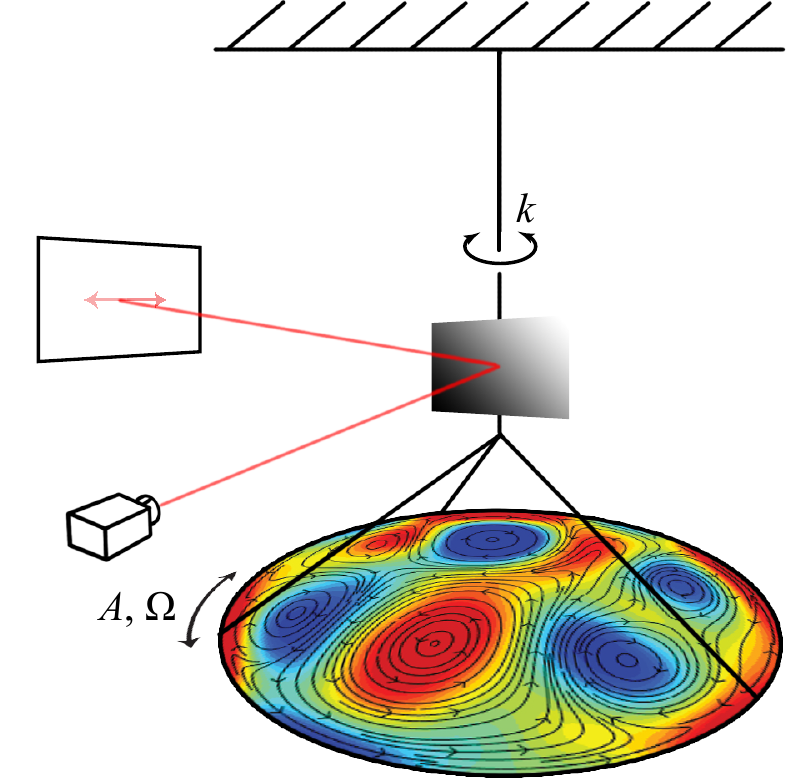}
 \caption{
Sketch of the proposed active-fluid analogue of the Andronikashvili experiment~\cite{Andronikashvili1948}. A rigid ring suspended on a torsional spring with the spring constant $k$ is filled with a thin-film active fluid that obeys Eq.~(\ref{e:eom}). We study the fluid-ring interaction in three scenarios: boundary held fixed ($k=\infty$, Fig.~\ref{fig02}), oscillatory motion of the ring induced by the torsional spring ($0<k<\infty$, Fig.~\ref{fig03}) and the response of the ring solely to the fluid stresses ($k=0$, Fig.~\ref{fig04}).
\label{fig01} 
}
\end{figure}

\par
To explore Stokes' second problem in the context of active fluids, we investigate a generalized Navier--Stokes model~\cite{1993BeNi_PhysD,2017SlomkaDunkel_PRF,2017SlomkaDunkel,2018Slomka_JFM,2018Mickelin_PRL} describing non-dilute active fluids that are subject to either oscillating boundary conditions or confined by a container that can respond freely to the internal fluid stresses. Inspired by recent experiments~\cite{2007SoEtAl,2007Ar}, we will specifically consider free-standing thin liquid films enclosed by a ring-shaped container of radius $R$ attached to a torsional spring of stiffness $k$ (Fig.~\ref{fig01}).  For containers periodically forced by a sufficiently stiff spring, our simulations predict an activity-induced reduction of the fluid inertia due to a lowered bulk viscosity~\cite{2015LoGaDoAuCle,ryan2011viscosity,2009SokolovAranson} and, hence, an decrease in the oscillation period. The experimental setup proposed in Fig.~\ref{fig01} is the active-fluid analogue  of the Andronikashvili experiment~\cite{Andronikashvili1948} used to measure the rotational oscillation frequency of a container filled with liquid helium. In the quantum case, decreasing the temperature leads to an increase in the ratio of the superfluid helium relative to the normal fluid phase. Since the superfluid phase decouples from the container dynamics, a decrease of temperature effectively reduces the oscillator mass~\cite{reppy2010nonsuperfluid,beamish2010supersolidity}, causing a measurable increase in the angular oscillation frequency. Our numerical results in Sec.~\ref{s:stokes} predict that active stresses can have a qualitatively similar effect, since topological defects in the bulk stress field can effectively decouple the bulk flow dynamics from the container.

\par
In the limit of a soft torsional spring ($k\to 0$), the same setup could be used to extract work from the collective microbial dynamics in an active fluid. Two recent experiments on bacterial~\cite{2016Wioland_RaceTracks} and microtubule~\cite{2017Wueaal} suspensions under channel confinement showed that active liquids can spontaneously achieve persistent circulation by exerting net forces on boundaries~\cite{2012Woodhouse,2017SlomkaDunkel_PRF,Theillard2017,thampi2016active}. Such non-equilibrium force generation raises interesting questions as to the combined dynamics of  isolated active-fluid--container systems~\cite{furthauer2012taylor}, implying a natural extension of the classic SPs. Whereas for passive fluids viscous friction eventually suppresses any container motion, active fluids can continually transform chemical into kinetic energy. This suggests that, under suitable conditions, mesoscopic bulk fluid vortices arising from collective microbial swimming could induce macroscopic fluctuations in the container's angular momentum, realizing an approximate non-equilibrium analogue of the Einstein--de Haas (EdH) effect~\cite{1915EinsteindeHaas}. In this crude analogy, the angular momenta of the bacterial vortices assume the role of the magnetic spin degrees of freedom, whose collective dynamics induces a measurable angular net motion of the macroscopic sample.  Our analysis in Sec.~\ref{s:EdH} for freely suspended containers  ($k\to 0$) driven by active fluid stresses indeed predicts that large resonant angular momentum fluctuations and, hence, work extraction can be achieved by tuning the container's diameter and the fluid-container mass ratio. 

\section{Model and numerical methods}

The subsequent analysis is based on a phenomenological higher-order stress model for actively driven solvent flow introduced in Refs.~\cite{2015SlomkaDunkel,2017SlomkaDunkel_PRF}. The energy transport characteristics of the resulting generalized Navier-Stokes equations (Sec.~\ref{s:GNS})  on periodic 2D and 3D periodic domains were characterized in earlier work~\cite{2017SlomkaDunkel,2018Slomka_JFM,2018Mickelin_PRL}. Here, we extend these studies to circular domains with stationary and explicitly time-dependent boundary conditions by making use of the recently developed double Fourier sphere spectral method~\cite{Wilber2017} (Sef.~\ref{s:numerics}). We note that conceptually similar higher-order PDEs have been successfully applied recently in the context of ionic liquids~\cite{2011Bazant,2012Storey}. Furthermore, closely related higher-order Navier--Stokes models have also been studied previously in the context of soft-mode turbulence, seismic waves~\cite{1993BeNi_PhysD,1996Tribelsky_PRL,PhysRevE.77.035202} and magneto-hydrodynamic turbulence\footnote{G. M. Vasil and M. G. P. Cassell, in preparation.}, so that the results below may have implications for these systems as well.

\subsection{Generalized Navier--Stokes equations for actively driven solvent flow}
\label{s:GNS}

We consider a passive incompressible solvent, such as water, driven by active stresses as generated by swimming bacteria~\cite{2012Sokolov}, ATP-powered microtubule bundles~\cite{2012Sanchez_Nature,2017Wueaal} or chemically or thermally propelled Janus particles~\cite{PhysRevLett.105.268302,2013Buttinoni_PRL}. The dynamics of the solvent velocity field $\bs v(t,\bs x)$ is described by the effective Navier--Stokes equations~\cite{2017SlomkaDunkel_PRF,2017SlomkaDunkel,2018Slomka_JFM,2018Mickelin_PRL}
\bse
\label{e:eom}
\be
\label{e:eoma}
\nabla\cdot\bs v&=&0, \\
\label{e:eomb}
\p_t \bs v+\bs v \cdot \nabla \bs v&=&-\nabla p+\nabla \cdot{\bs \sigma},
\ee
where   $p(t,\bs x)$ is the local pressure. The stress tensor $\bs \gs(t,\bs x)$ comprises passive and active contributions, representing the intrinsic solvent fluid viscosity and stresses exerted by the microswimmers on the fluid~\cite{2002Ra,2008SaintillanShelley,2013Marchetti_Review,2013Ravnik_PRL}. As shown recently~\cite{2017SlomkaDunkel},  a minimal linear extension of the usual Navier--Stokes for passive Newtonian fluids, 
\be
\label{e:stress}
\bs\gs=(\Gamma_0 -\Gamma_2 \nabla^2+\Gamma_4 \nabla^4)[\nabla \bs v+ (\nabla \bs v)^\top],
\ee
\ese
suffices to quantitatively reproduce experimentally measured bulk flow correlations in bacterial and microtubule suspensions~\cite{2004DoEtAl,2007Cisneros,2012Sanchez_Nature,2012Sokolov,Dunkel2013_PRL,Wioland2013_PRL}. The empirical fit parameters $\Gamma_{0}>0$, $\Gamma_{2}<0$ and $\Gamma_4>0$ determine  the most unstable mode corresponding to the characteristic vortex size \mbox{$\Lambda=\pi \sqrt{2\Gamma_4/(-\Gamma_2)}$}, the typical growth timescale $\tau$ and the bandwidth $\kappa$ of the unstable wavenumber range~\cite{2017SlomkaDunkel}
$$
\tau = \left[\f{\Gamma_2}{2\Gamma_4}\left(\Gamma_0-\f{\Gamma_2^2}{4\Gamma_4}\right)\right]^{-1},
\quad
\kappa =
\biggl(\f{-\Gamma_2}{\Gamma_4}-2\sqrt{\f{\Gamma_0}{\Gamma_4}}\biggr)^{1/2}.
$$
The typical vortex circulation speed is $U=2\pi\Lambda/\tau$. The bandwidth $\kappa$ controls the active fluid mixing and spectral energy transport from smaller to larger scales~\cite{2017SlomkaDunkel}. We choose $(\Lambda,\tau,\kappa)$ to characterize active flow structures, as these parameters can be directly inferred from experimental data~\cite{2017SlomkaDunkel}. Length and time will be measured in units of $\Gl$ and $\tau$ from now on. Equation~\eqref{e:eom} describes a non-dilute regime, in which collective active dynamics is dominant and advection may not be ignored~\emph{a priori} (Appendix~\ref{app:nonlinearities}). 
\par
Focusing on a planar disk domain of radius $R$, we can rewrite Eqs.~(\ref{e:eom}) in the vorticity-stream function form
\bse
\label{e:eomvsf}
\be
\label{e:eomvsfa}
\p_t \omega+\nabla\omega\wedge\nabla\psi
&=& 
\Gamma_0\nabla^2\omega-\Gamma_2\nabla^4\omega+\Gamma_4\nabla^6\omega, 
\qquad\\
\label{e:eomvsfb}
\nabla^2\psi&=&-\omega,
\ee
\ese
where the vorticity pseudo-scalar  $\omega=\nabla\wedge \bs v=\eps_{ij}\partial_iv_j$  is defined in terms of the 2D Levi--Civita tensor $\eps_{ij}$, and $\psi$ is the stream function. In polar coordinates $(r,\theta)$, one recovers the radial and azimutal velocity components from $v_r=({1}/{r})\p_\theta \psi$ and $v_\theta=-\p_r\psi$. An impermeable container wall imposes the radial boundary condition $v_r(t,R,\theta)=0$. The tangential component satisfies the no-slip condition $v_\theta(t,R,\theta)=V(t)$. We will consider three cases: a stationary boundary $V(t)=0$, periodic forcing  $V(t)\approx A\cos(\Go t)$ and freely suspended boundaries, where the fluid stresses induce a rigid body motion $V(t)$ of the container. Additionally, we fix soft higher-order boundary conditions $\nabla^2\omega(R,\theta)=\nabla^4\omega(R,\theta)=0$ throughout, which have been shown previously to reproduce the experimentally observed bulk flow dynamics and viscosity reduction in rectangular shear geometries~\cite{2017SlomkaDunkel_PRF}.

\begin{figure}[b]
\centering
\includegraphics[width=0.6\textwidth]{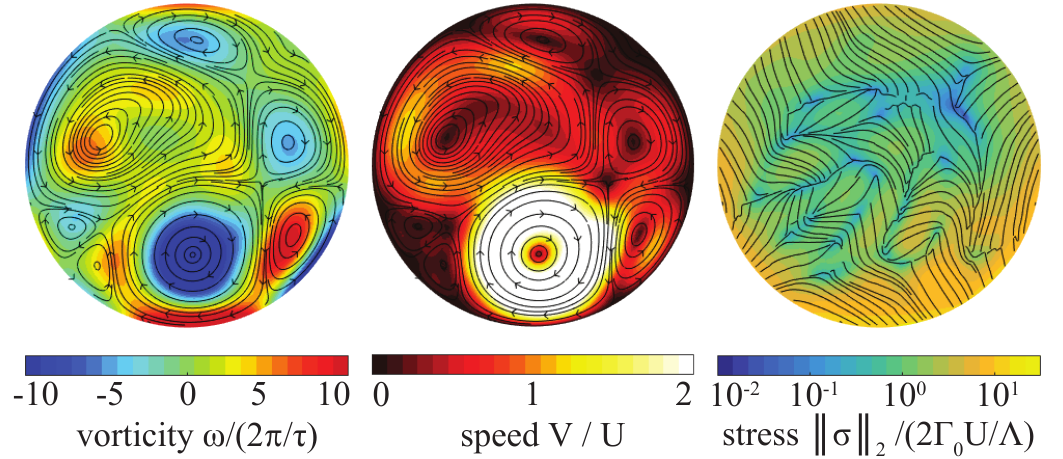}
 \caption{
 Typical flow and stress fields for an active fluid with vortex size $\Lambda$ and wide vortex-size distribution $\kappa_\tn{w}=1.5/\Lambda$, confined to a planar disk geometry (radius $R=2.67\Lambda$) with boundary held fixed. The presence of the stress-free defects allows the stress director field to develop complex configurations, enabling a nontrivial response to time-dependent boundary conditions, see Figs.~\ref{fig03} and~\ref{fig04}.
\label{fig02} 
}
\end{figure}

\subsection{Numerical method and stationary boundary}
\label{s:numerics}
To solve Eq.~(\ref{e:eomvsf}) numerically with spectral accuracy, we implemented a recently developed disk analogue of the 
double Fourier sphere method~\cite{Wilber2017}. The underlying algorithm uses a polar coordinate representation while avoiding the introduction of an 
artificial boundary at the origin. We combined this method with a third-order IMEX time-stepping scheme, which decouples the system of PDEs~(\ref{e:eomvsf}) and treats the nonlinear advection term explicitly.  Spatial differential operators were discretized using 
the Fourier spectral method in $\theta$ and the ultraspherical spectral method in~$r$~\cite{OlverTownsend2013}. This procedure
generates a sparse spectrally-accurate discretization that can be solved in a cost of $\mathcal{O}(n^2 \log n)$ operations per time step, 
where $n$ is the number of Fourier--Chebyshev modes employed in $\theta$ and $r$.  To avoid aliasing errors,  the 3/2-rule~\cite{SpecMethodsFD} was used to evaluate the advection term. Additional mode-filtering prevents unphysical oscillations in the solution (see~\cite{TorresCoutsias1999} for details). The no-slip boundary conditions were enforced via integral conditions on the vorticity field~\cite{Quartapelle1981}.

\begin{figure*}[b!]
\centering
\includegraphics[width=0.95\textwidth]{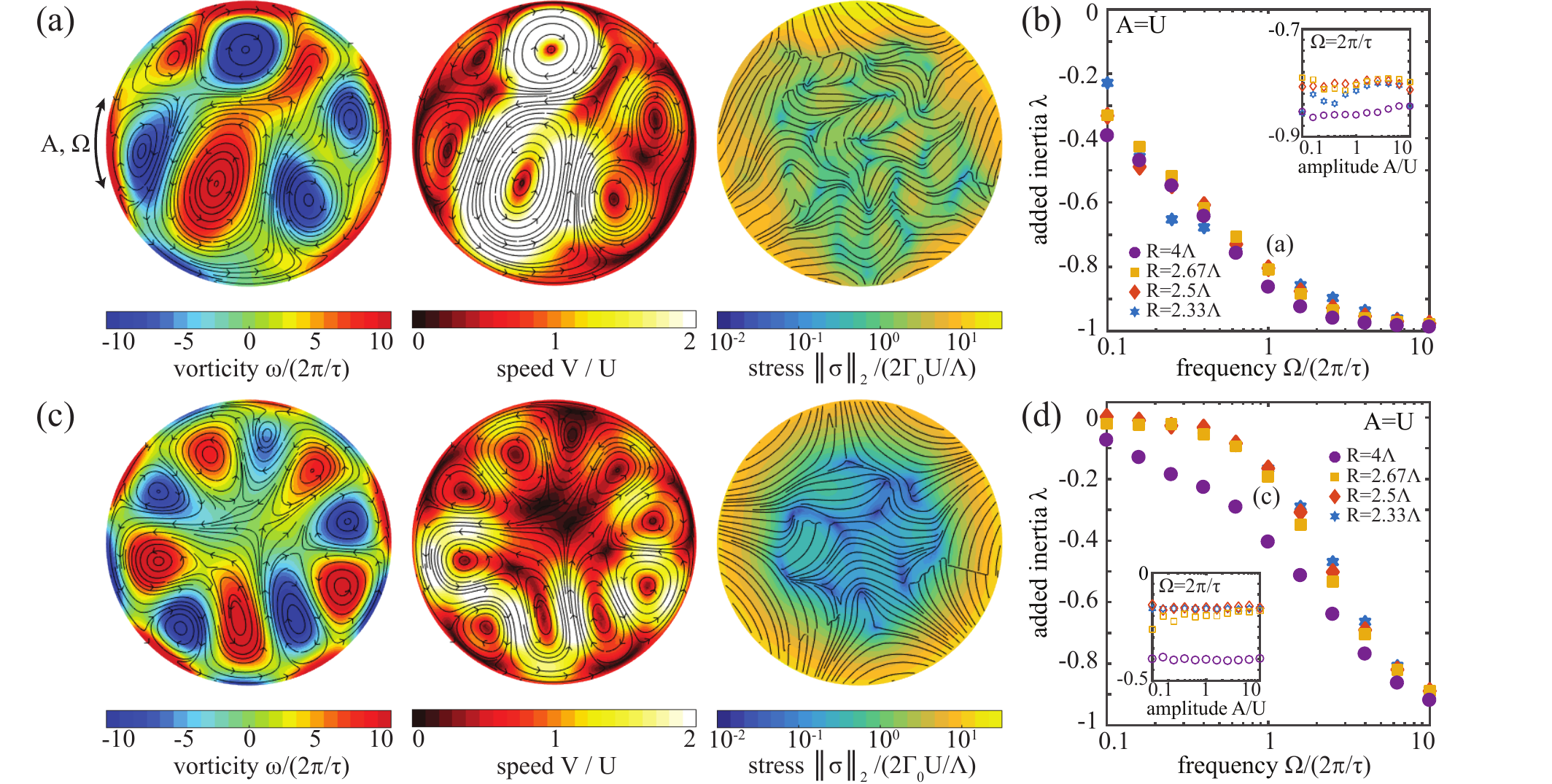}
 \caption{
Stokes' second problem for a ring-shaped container pendulum coupled to a torsional spring. Response of active fluids with wide (a,b: $\kappa_{\tn{w}}=1.5/\Lambda$) and small (c,d: $\kappa_{\tn{s}}=0.63/\Lambda$) vortex-size distributions to oscillatory boundary conditions (see Movies 2 and~3 and Fig.~\ref{figSI02}). The boundary speed is sinusoidal with amplitude $A$ and angular frequency $\Omega$. (b,d) The activity-induced relative change $\lambda$  in the effective inertia experienced by the ring pendulum. Negative values of $\lambda$ imply that the pendulum oscillates at higher frequency in an active fluid than in a passive fluid.
\label{fig03} }
\end{figure*}
\par
We first solved Eqs.~(\ref{e:eomvsf}) for the fixed boundary conditions, $v_{\theta}(t,R,\theta)\equiv 0$, which corresponds to the limit of infinite spring constant $k\to\infty$ in Fig.~\ref{fig01}. In the case of a relatively wide bandwidth $\kappa_\tn{w}=1.5/\Lambda$,  the active flow spontaneously forms vortices spanning a range of different diameters in the vicinity of the preferred value~$\Lambda$  (Fig.~\ref{fig02} and Movie 1), in agreement with recent simulations~\cite{Theillard2017} of multi-field models.   The traceless nematic stress tensor field $\bs\sigma$ defined in Eq.~\eqref{e:stress} is uniquely  characterized by its largest eigenvalue $\|\bs{\sigma}\|_2$ and the director field of the corresponding eigenvector. We generally find that the stress director fields develops locally ordered domains, which are punctured and separated by stress-free topological defects (Fig.~\ref{fig02} right).  As we shall see below,  the defects facilitate activity-induced reduction of the fluid inertia, when the container is periodically forced.

\section{Stokes' second problem and activity-induced reduction of added mass}
\label{s:stokes}
To connect with Stokes' second problem, we next consider the motion of a ring pendulum consisting of a circular container coupled to a torsional spring with a finite stiffness  constant $0<k<\infty$ in Fig.~\ref{fig01}. The torque exerted by an active fluid of mass $m_\mathrm{f}$ on the ring is (Appendix~\ref{app:ang_mom_eq})
\be
\mathcal{T}=-\f{m_\mathrm{f}}{\pi}\int_0^{2\pi}d\theta\, \sigma_{r\theta}(t,R,\theta),
\ee
with $\sigma_{r\theta}$ the normal-tangential component of the stress tensor (\ref{e:stress}) in polar coordinates. 
Our simulations show that relation between $\mathcal{T}$ and the angular speed of the ring, $\dot\phi=v_{\theta}/R$, is dominated by linear response (Appendix~\ref{app:stokes_linear_res}), 
\be
\label{e:linear_response}
\mathcal{T}=-I_\tn{f}\ddot\phi-\gamma\dot\phi,
\ee
with the inertial and dissipative parameters $I_\tn{f}$ and $\gamma$ depending on the driving frequency, geometry and fluid parameters.  
For passive fluids at low Reynolds number, Eqs.~(\ref{e:linear_response}) holds exactly, and $I_\tn{f}$ and $\gamma$ can be calculated for simple geometries, owing to the linearity of the Stokes' equations~\cite{landau2013fluid}. For our active fluid model, we can determine $I_\tn{f}$ and $\gamma$ directly from the numerically measured power spectral densities (Appendix~\ref{app:stokes_linear_res}).  To find out how activity affects the pendulum frequency, we follow Stokes' original argument~\cite{stokes1851effect} and balance $\mathcal{T}$ with the torque exerted by the torsional spring of stiffness $k$,  which yields
\be
(I_\tn{c}+I_\tn{f})\ddot\phi+\gamma\dot\phi+k\phi=0,
\ee
where $I_\tn{c}= m_\tn{c} R^2$ is moment of inertia of a ring of mass~$m_\tn{c}$. Since the effect of $\gamma$ is generally quite small (Appendix~\ref{app:stokes_diss_res}), we find 
that to leading order $v_{\theta}(t,R,\theta)=A\cos{(\Omega t)}$, where $\Omega=[k/(I_\tn{c}+I_\tn{f})]^{1/2}$. For passive fluids, this is exactly the result obtained by Stokes', who concluded that the added fluid inertia $I_\tn{f}$ reduces a pendulum's frequency~$\Omega$. Moreover, by expressing $I_\tn{f}$ in terms of viscosity,  he was then able to explain several puzzling experiments~\cite{stokes1851effect}. For parameters relevant to microbial experiments,  a passive fluid essentially behaves as a rigid body since the penetration depth $\sqrt{2\Gamma_0/\Omega}$ is much larger than the container radius $R$ (Appendix~\ref{app:stokes_passive}). In this case, the moment of inertia of the passive fluid equals that of a solid disk, $I_\tn{f,p}=\f{1}{2}m_\tn{f}R^2$. Using $I_\tn{f,p}$ as a natural reference point, we express the effective inertia of an active fluid as $I_\tn{f,a}=(1+\lambda)I_\tn{f,p}$, where  $\lambda$ is the relative added inertia due to activity. To explore how confinement geometry, driving protocol and active fluid properties affect $\lambda$,  we varied systematically  the amplitude~$A$,  the oscillation frequency~$\Omega$, and the container radius $R$ in our simulations, comparing active fluids with wide  ($\kappa_{\tn{w}}=1.5/\Lambda$) and small ($\kappa_{\tn{s}}=0.63/\Lambda$)  energy injection bandwidths, respectively~(Fig.~\ref{figSI02}). Interestingly, we find that for both values of $\kappa$, the added inertia is negative, $\lambda<0$, across a wide range of driving frequencies $\Omega$ and amplitudes $A$ [Figs.~\ref{fig03}(b,d)]. This implies that the fluid activity effectively reduces the amount of inertia transferred to the pendulum, and hence increases the oscillation frequency compared with a passive fluid. At high frequencies $\Omega\gg 2\pi/\tau$, which can be achieved by using sufficiently stiff springs, $\lambda\approx -1$ implying that the pendulum does not acquire additional inertia and oscillates as if placed in a vacuum. In this regime, the bulk flow effectively decouples from the boundary due to the presence of defects in the stress field.

\section{Work extraction from geometrically quantized active fluctuations} 
\label{s:EdH}

Mimicking the classical EdH-setup, we now consider a container-fluid system isolated from external forces or torques ($k\to 0$), so that the container responds solely to the stresses generated by the enclosed fluid. In passive fluids, viscosity dissipates energy and such a system will eventually converge to a state of rest or rigid rotation if it had nonzero initial angular momentum. By contrast, active fluids are continuously supplied with kinetic energy through conversion of chemical energy and may thus induce a permanent dynamic response of the container. Focusing as before on a thin rigid ring-shaped container governed by Newton's second law, the angular dynamics of the ring is determined by (Appendix~\ref{app:ang_mom_eq})
\be
\label{e:angmomODE}
\ddot{\phi}=-\f{\alpha}{\pi}\int_0^{2\pi}d\theta\, \sigma_{r\theta}(t,R,\theta),
\ee
where $\alpha=m_\tn{f}/m_\tn{c}$ is the ratio of total fluid mass and ring mass. We solve Eqs.~(\ref{e:eomvsf}) and~\eqref{e:angmomODE} simultaneously using $V(t)=R\dot{\phi}$ as boundary condition for Eqs.~(\ref{e:eomvsf}).  

\par
To interpret the simulation results, we note that the characteristic length and time scales $\Lambda$ and $\tau$ of an active fluid give rise to a natural unit of angular momentum. Regarding a single vortex as a thin rigid disk of radius $\Lambda/2$ rotating at the constant angular speed $2\pi /\tau$, one finds the characteristic kinematic angular momentum $L_{\tn{v}}=\pi^2 \Lambda^4 /(16\tau)$. A planar disk of radius $R$ can carry about $N_{\tn{v}}=(2R/\Lambda)^2$ vortices, so it is convenient to introduce the normalization factor $\ell=\sqrt{N_{\tn{v}}}L_{\tn{v}}$. Adopting $\ell$ as basic unit, one would expect specific angular momentum fluctuations of order one if $N_\tn{v}$ vortices contributed randomly in uncorrelated manner. Larger fluctuations indicate correlated collective angular momentum transfer between vortices and the boundary.
\par
\begin{figure}[t!]
\centering
\includegraphics[width=0.5\textwidth]{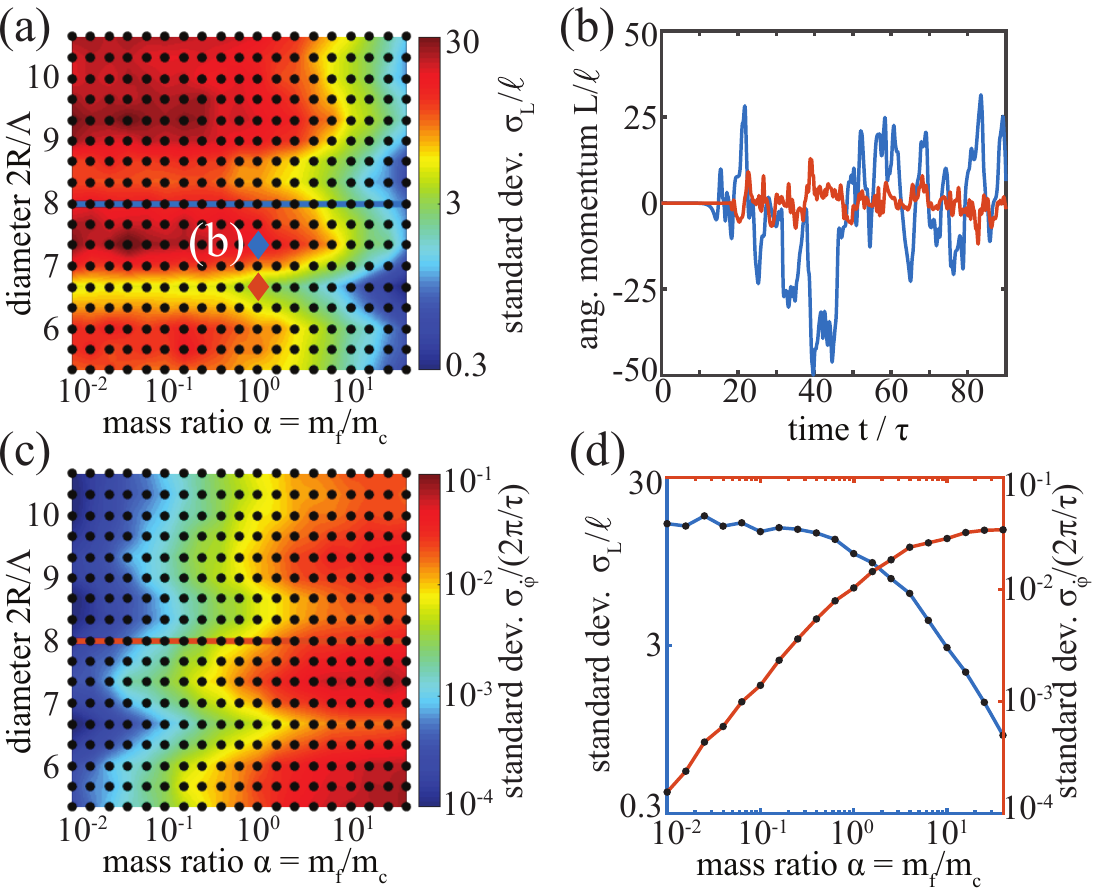}
 \caption{
Geometrically quantized fluctuations in an active fluid with narrow vortex-size distribution $\kappa_{\tn{S}}=0.63/\Lambda$. When isolated, the fluid can significantly shake the enclosing container, a thin rigid ring of radius $R$.
(a) The standard deviation $\sigma_L$ of the container angular momentum $L$ depends on the radius $R$ and the fluid-to-ring mass ratio $\alpha$. The fluctuations $\sigma_L$  are independent of $\alpha$ for heavy containers ($\alpha \ll 1$) but start to decrease monotonically with $\alpha$ when the containers become light ($\alpha \gg 1$). As $R$ varies, the fluctuations oscillate with the period set by the characteristic vortex scale $\Lambda$ (see also Fig.~\ref{figSI06}). Black dots represent 323 simulated parameter pairs, the color code shows linear interpolation. 
(b) Representative time series of the container angular momentum for two different radii $R=3.67\Lambda$ (Movie 4) and $R=3.33\Lambda$ (Movie 5) and fixed mass ratio $\alpha=1$.
(c) The standard deviation $\sigma_{\dot{\phi}}\sim \alpha \sigma_L$ of the container angular speed $\dot{\phi}$.
(d) Horizontal cuts through (a) and (c) at constant radius $R=4\Lambda$. In particular, to maximize both the angular momentum and velocity fluctuations, the fluid mass should match the container mass ($\alpha\sim 1$).
}
\label{fig04} 
\end{figure}

Focusing on an active fluid with narrow vortex-size distribution ($\kappa_\tn{s}=0.63/\Lambda$), we performed parameter scans to determine how the standard deviations $\sigma_L$ and $\sigma_{\dot{\phi}}$ of  the ring's angular momentum $L$ and angular speed $\dot{\phi}$ depend on the ring radius $R$ and fluid-to-ring mass ratio $\alpha$. Our simulations show that for a heavy container ($\alpha \ll 1$), the fluctuations $\sigma_L$ are approximately independent of $\alpha$, in which case their magnitude is the same as if the boundary was held fixed (cf.~Fig~\ref{fig02}). Once the container becomes lighter ($\alpha \sim 1$),  the fluctuations start to decrease, with the decay rate approaching $1/\alpha$ for very light containers ($\alpha \gg 1$) [Figs~\ref{fig04}(a,d)]. Similarly, the angular velocity fluctuations $\sigma_{\dot{\phi}}\sim \alpha\sigma_{L}$  are independent of $\alpha$ for light containers, but increase linearly for heavy containers. In particular, $\sigma_{\dot{\phi}}$ vanishes as $\alpha\to 0$, implying that the container becomes stationary as its mass becomes very large, as expected  [Figs~\ref{fig04}(c,d)]. We also conclude that to maximize the angular velocity fluctuations $\sigma_{\dot{\phi}}$ without significantly reducing the angular momentum transfer to the boundary the fluid mass should match the container mass  ($\alpha\sim 1$). Strikingly, we find that the fluctuations oscillate as a function of $R$, with the period set by the vortex size $\Lambda$ [Figs~\ref{fig04}(a,c); Movie 4 and 5; Fig.~\ref{figSI06}]. This result corroborates the idea~\cite{Lee2017PNAS} that non-monotonic energy spectra, which the dynamical system~(\ref{e:eom}) develops~\cite{2017SlomkaDunkel}, generically result in oscillatory forces on boundaries~\cite{Lee2017PNAS}. For optimal combinations of $(\alpha,R)$,  $\sigma_L$ can be more than an order of magnitude larger than $\ell$, indicating that the bulk vortices transfer angular momentum to the container collectively [Fig.~\ref{fig04}(b)]. Such large quantized non-equilibrium fluctuations offer a novel way of extracting work from active suspensions (e.g., with fluctuation-driven microelectrical alternators), complementing recently proposed ratchet-based devices~\cite{Sokolov19012010,di2010bacterial}. For example, at the peak of bacterial activity we may take $\tau\sim2$s and $\Lambda\sim50\mu\tn{m}$~\cite{Dunkel2013_PRL,2017SlomkaDunkel}. At the mass ratio $\alpha\sim 1$ and the container radius $R=200\mu\tn{m}$, Fig.~\ref{fig04}(d) gives $\sigma_{\dot{\phi}}\sim 0.02(2\pi/\tau)\sim 0.06\,\tn{rad}/\tn{s}$, comparable with the rotation rates of bacteria-powered microscopic gears of similar size reported in~\cite{Sokolov19012010}.

\section{Conclusions}
Recent experiments~\cite{2016Wioland_RaceTracks,Wioland2013_PRL,2016Wioland_NPhys,2017Wueaal} have successfully utilized the interplay between characteristic flow pattern scales in active turbulence and confinement geometry to rectify and stabilize collective dynamics in natural and synthetic microswimmer suspensions. The above analysis extends these ideas to the time domain to achieve dynamic control, similar in spirit to actuation-controlled classical turbulence~\cite{tardu2017wall}. Our two main predictions about activity-induced reduction of fluid inertia and geometrically quantized large fluctuations for a freely suspended container-fluid system should be testable with recently developed experimental techniques~\cite{2007SoEtAl,2007Ar}.


\appendix

\section{Nondimensionalization}
\label{app:nondim}
For numerical simulations, we non-dimensionalize the equations of motion~(\ref{e:eomvsf}) by rescaling according to
\be
t'=T_0t,\quad x_i'= R x_i, \quad \omega'= \omega_0 \omega, \quad \psi'=\psi_0 \psi,
\ee
which gives, after dropping the primes,
\be
\p_{t}\omega+T_0\f{\psi_0}{R^2}(\p_{ y}\psi)\p_{ x}\omega-T_0\f{\psi_0}{R^2}(\p_{x} \psi)\p_{ y}\omega&=&\f{T_0 \Gamma_0}{R^2}\Big(\nabla^2 \omega -\f{{\Gamma}_2}{\Gamma_0 R^2}\nabla^4  \omega+\f{\Gamma_4}{\Gamma_0 R^4}\nabla^6 \omega\Big), \\\notag
\f{\psi_0}{R^2}\nabla^2\psi&=&-\omega_0\omega.
\ee
We set $\psi_0=\f{R^2}{T_0}$ and $\omega_0=\f{1}{T_0}$ and $T_0=\f{R^2}{\Gamma_0}$, which leads to
\be
\p_{t}\omega+(\p_{ y}\psi)\p_{ x}\omega-(\p_{x} \psi)\p_{ y}\omega&=&\nabla^2 \omega -\gamma_2\nabla^4  \omega+\gamma_4\nabla^6 \omega, \\\notag
\nabla^2\psi&=&-\omega,
\ee
where $\gamma_2=\f{\Gamma_2}{\Gamma_0R^2}$ and $\gamma_4=\f{\Gamma_4}{\Gamma_0 R^4}$. 

\section{Advective nonlinearities}
\label{app:nonlinearities}
The applicability of model~(\ref{e:eom}) is restricted to active fluids driven by intermediate concentrations of self-propelling ingredients (bacteria, microtubules), for which elastic energy
transfer within the active component can be neglected, but the advective nonlinearities must be retained. This is because at such concentrations collective behavior persists, implying that the Reynolds number is not small~\cite{2017SlomkaDunkel_PRF}, in contrast to flows typical for dilute suspensions. Such intermediate concentrations exist, see for example~\cite{2017Wueaal} where at volume fractions as low as 0.05\% a microtubule network drives bulk turbulent flows. In this section we give an alternative argument showing that inertial nonlinearities cannot \emph{a priori} be neglected in any continuum model aiming to describe statistically stationary active turbulence.
\par
Any continuum description of the solvent flow in active fluids must in principle respect momentum conservation as expressed by the Cauchy equations
\be
\label{e:generic_active_theory}
\p_t \bs v+\bs v \cdot \nabla \bs v&=&-\nabla p+\nabla \cdot{\bs \sigma}.
\ee
In active fluids, the stress tensor $\bs\sigma$ can be usually split into two competing parts
\be
\bs\sigma=\bs\sigma_\tn{passive}+\bs\sigma_\tn{active},
\ee
where the active part represents stresses due to microswimmers, microtubule extensile networks, etc., depending on the particular type of active driving, while the passive part arises due to dissipative mechanisms such as solvent viscosity. Assuming incompressible flow, the corresponding fluid kinetic energy equation is given by
\be
\p_t \Big(\f{1}{2}|\bs v|^2\Big)+\bs v \cdot \nabla \Big(\f{1}{2}|\bs v|^2\Big)&=&-\nabla \cdot(p\bs v)+\bs v\cdot\nabla \cdot{\bs \sigma_\tn{passive}}+\bs v\cdot\nabla \cdot{\bs \sigma_\tn{active}}.
\ee
The left-hand side is the advective derivative of the kinetic energy density. The right-hand side is a scalar field that represents contributions to the rate of change of the kinetic energy along the flow due to work done by pressure, passive stresses and active stresses. In the statistically stationary state, the global kinetic energy fluctuates around some mean value. In this regime, the rate of change of kinetic energy cannot be everywhere negative at all times, for this would render the whole system dissipative and the solvent would stop moving. Similarly, it cannot be everywhere positive at all times, for this would render the system unstable. By continuity, there should be domains where passive and active parts cancel, as suggested by the recent observation of bacterial \lq superfluids\rq~\cite{2015LoGaDoAuCle,PhysRevLett.118.018003}.  In these domains, the advective nonlinearities may play a central role for energy transport. Furthermore, in dilute bacterial suspensions, it is known that the Stokes' approximation holds to an excellent degree by the usual low Reynolds number argument based on a single microswimmer in infinite bath~\cite{1977Pu}. However, higher microswimmer concentrations are characterized by steric and hydrodynamic interactions that lead to coherent motion. In this case, the Reynolds number associated with these coherent structures are not necessarily small~\cite{2017SlomkaDunkel_PRF}. This is particularly true  for non-dilute suspensions of larger faster-swimming cells, such as sperm.

\vspace{0.5cm}
\section{Stokes' second problem: driving protocol}
\label{app:stokes_driving_protocol}
\begin{figure}[h!]
\centering
\includegraphics[width=0.5\textwidth]{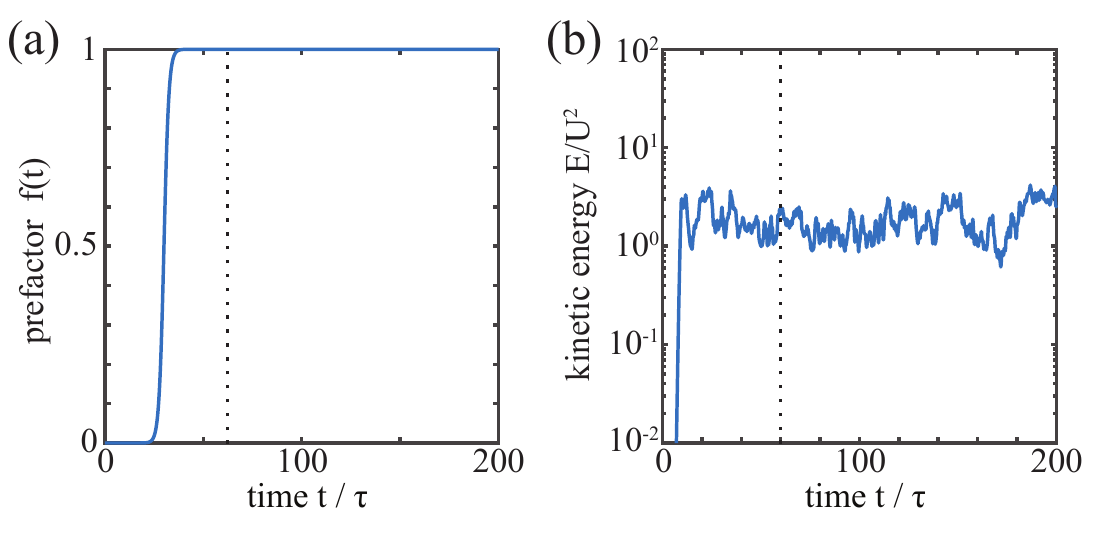}
 \caption{
Stokes' second problem: Driving protocol. (a) The driving amplitude, cf.~Eq.~(\ref{e:stokes_bc}), increases according to the prefactor $f(t)$ defined in Eq.~(\ref{e:prefactor}). (b) Mean kinetic energy time series for the simulation shown in Fig.~\ref{fig03}(b) shows that the system relaxes well before the start time of the temporal averaging periods for the mean power $\langle P \rangle$ (vertical dashed lines).}
\label{figSI01} 
\end{figure}

We describe in detail the driving protocol for the active Stokes' second problem. At $t=0$, we initiate the simulations with both the boundary and fluid at rest plus a small random perturbing flow $\delta \bs v$ ($\|\delta \bs v\|_1/U\ll 1$, where $U$ is the characteristic speed of the turbulent patterns). We then turn on the periodic driving by applying the no-slip boundary condition
\be
\label{e:stokes_bc}
v_\theta(t,R,\theta)=f(t)A\cos (\Omega t),
\ee
where
\be
\label{e:prefactor}
f(t)=\f{1}{2}\Big[1+\tn{tanh}\Big(\f{t-30\tau}{2.5\tau}\Big)\Big],
\ee
and $\tau$ is the characteristic time scale of the active flow patterns as defined in the Introduction. Thus, for the time interval $\sim 30\tau$ the boundary remains stationary. During that time, the bulk flow relaxes and active turbulence develops. At about $\sim 30\tau$, the periodic driving sets in and the boundary condition~(\ref{e:stokes_bc}) quickly approaches $v_\theta(t,R,\theta)=A\cos (\Omega t)$. Calculation of the mean power $\langle P \rangle$ presented in Fig.~\ref{fig03} occurs during the time interval $[60\tau, 200\tau]$, long after the relaxation, see Fig.~\ref{figSI01}.

\begin{figure}[h!]
\centering
\includegraphics[width=1.0\textwidth]{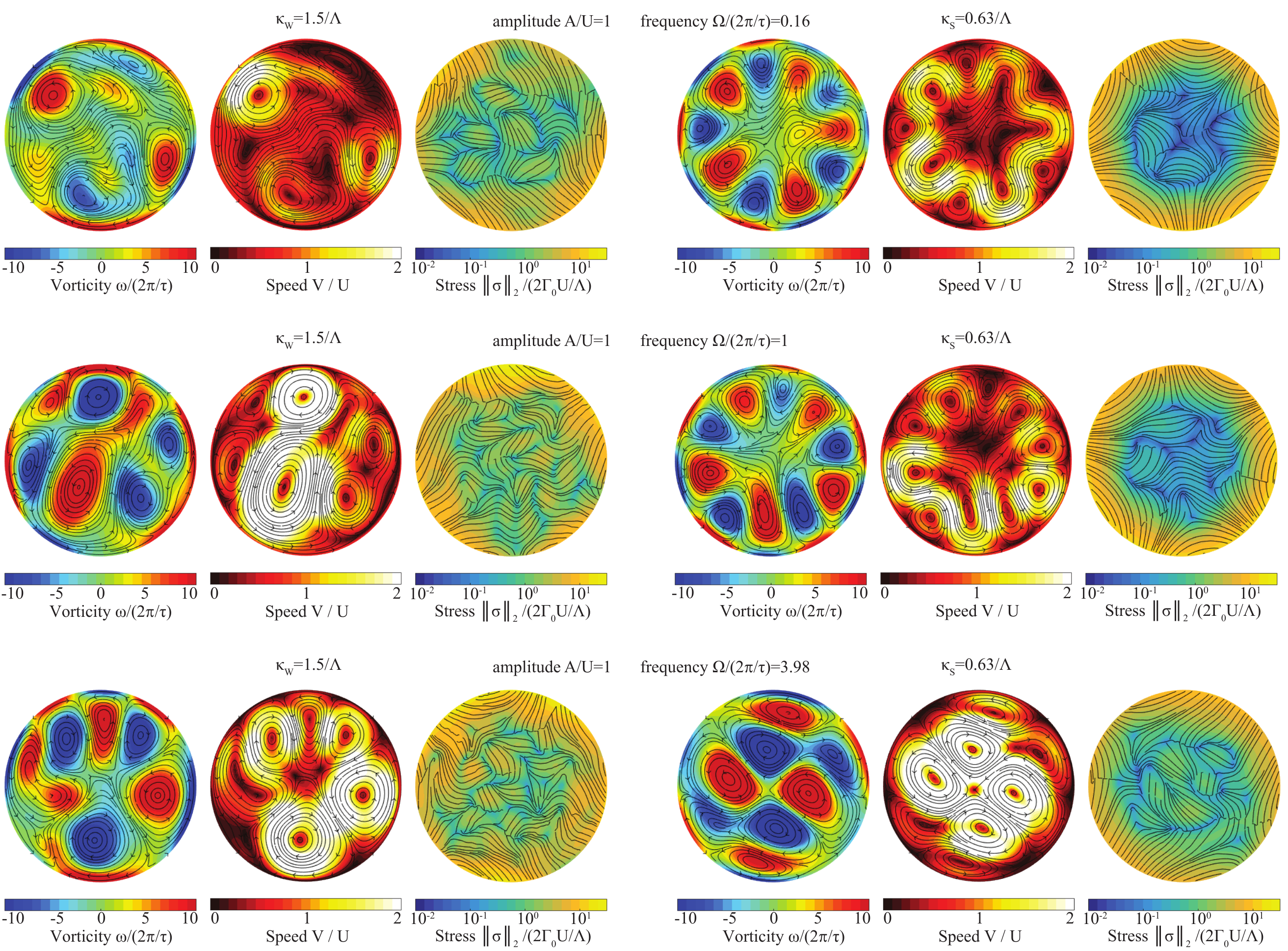}
 \caption{
Stokes' second problem: Additional examples. The middle row is the same as Figs.~\ref{fig03}(b,d). For narrow vortex-size distribution $\kappa_\tn{s}$ (right column), it takes higher driving frequency $\Omega$ to disrupt the structure of the stress field, a core filled with low stress defects and half-loops of the stress director field near the boundary.
}
\label{figSI02} 
\end{figure}

\section{Stokes' second problem: Linear relation between torque and angular speed}
\label{app:stokes_linear_res}
In this section we quantify how accurately the formula~(\ref{e:linear_response}), which relates the fluid-induced torque $\mathcal{T}(t)$ on the container to the container angular speed $\dot\phi(t)$, describes the response of an active fluid subject to oscillatory boundary conditions. This formula approximately holds if the power spectral density (PSD) $|\mathcal{T}(\omega)|^2$ of the time-series $\mathcal{T}(t)$ is concentrated at the driving frequency $\Omega$. To see this, we follow the usual argument, see for example~\cite{landau2013fluid}. We write the container angular speed as $\dot\phi=\phi_0\Re(e^{i\Omega t})$, where $\phi_0=A/R$. If the PSD $|\mathcal{T}(\omega)|^2$ is concentrated at $\Omega$, then
\be
\label{eq:approx_lin_res}
\mathcal{T}(t)
\approx
\Re\left\{\mathcal{T}(\Omega)e^{i\Omega t}]=\Re\{[\mathcal{T}_\tn{r}(\Omega)+i\mathcal{T}_\tn{i}(\Omega)]e^{i\Omega t}\right\}
=
\Re\left\{\mathcal{T}_\tn{r}(\Omega)e^{i\Omega t}+\f{\mathcal{T}_\tn{i}(\Omega)}{\Omega} \f{d}{dt}e^{i\Omega t} \right\}
=
\f{\mathcal{T}_\tn{r}(\Omega)}{\phi_0}\dot\phi(t)+\f{\mathcal{T}_\tn{i}(\Omega)}{\phi_0\Omega}\ddot\phi(t).
\ee
Setting $I_\tn{f}=-[\mathcal{T}_\tn{r}(\Omega)]/\phi_0$ and $\gamma=-[\mathcal{T}_\tn{i}(\Omega)]/[\phi_0\Omega]$, we obtain Eq.~(\ref{e:linear_response}).
\par
To verify that the PSD $|\mathcal{T}(\omega)|^2$ of the time-series $\mathcal{T}(t)$ is concentrated at the driving frequency $\Omega$, we performed spectral analysis of the steady-state part of $\mathcal{T}(t)$. Let $\mathcal{T}_n$ be the discrete time series obtained in simulations, where $n$ denotes the time step. The number of time steps is always taken large enough to ensure that the physical time interval is at least two orders of magnitudes greater than the larger of the two quantities: the characteristic pattern formation scale $\tau$ or the driving period $T=2\pi/\Omega$. The time series itself is obtained by integrating the stress-tensor over the container according to the Eq.~(\ref{eq:torque_small_element}). We apply the Discrete Fourier Transform to $\mathcal{T}_n$ to obtain the discrete PSD $|\mathcal{T}_\omega|^2$.
\par
\begin{figure}[t!]
\centering
\includegraphics[width=0.95\textwidth]{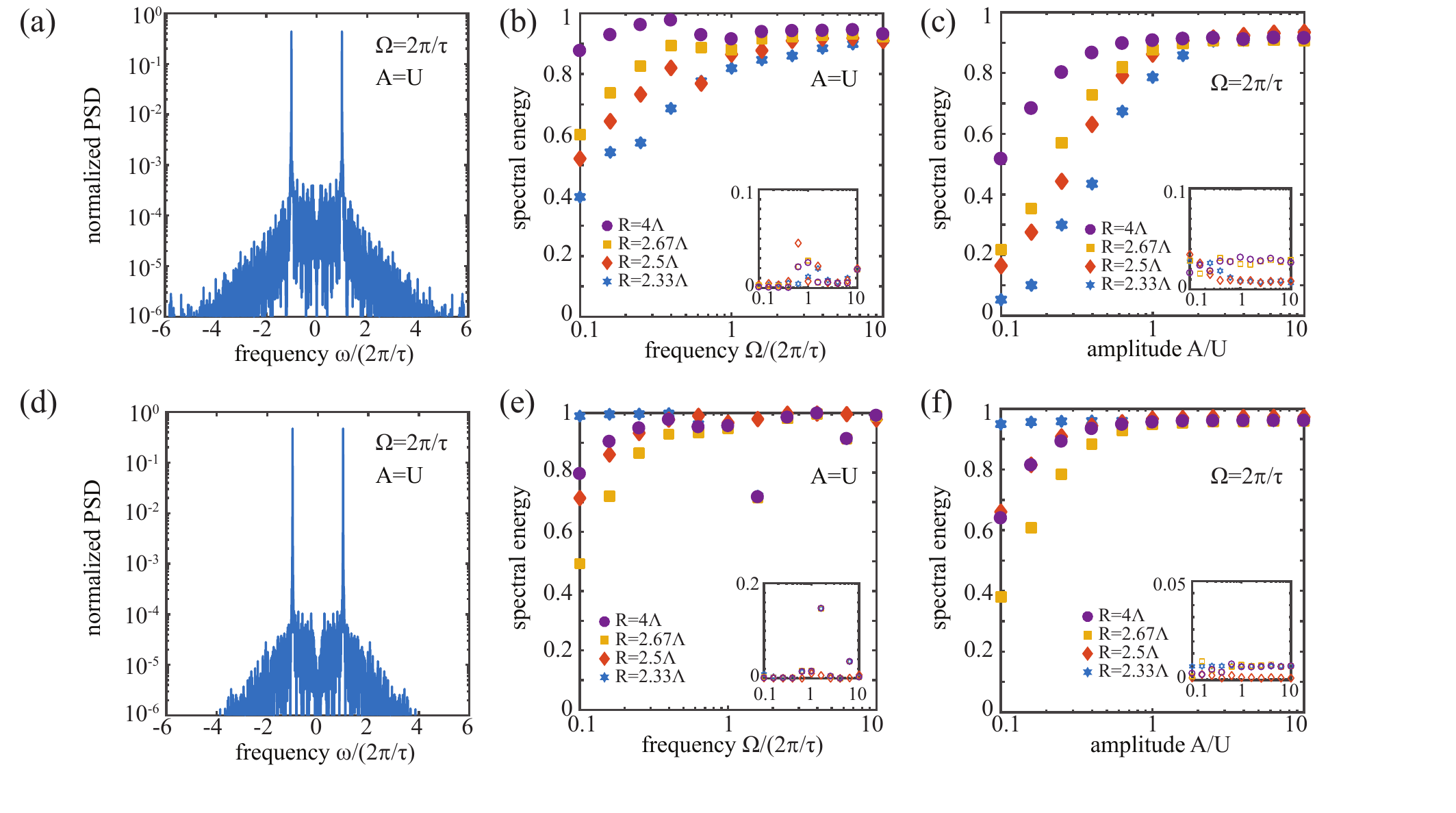}
 \caption{
The linear relation~(\ref{e:linear_response}) between the torque $\mathcal{T}(t)$ and angular speed $\dot\phi(t)$ approximately holds for active fluids. The formula becomes very accurate as the driving frequency $\Omega$ and amplitude $A$ become larger than the corresponding active fluid characteristic pattern formation parameters $2\pi/\tau$ and $U$, respectively. (a,d) Normalized power spectral density $|\mathcal{T}_\omega|^2/\sum_{\omega'} |\mathcal{T}_{\omega'}|^2$ of the (discrete) steady-state time series $\mathcal{T}_n$ for the two simulations shown in Figs.~\ref{fig03}(b,d). The complex amplitudes $\mathcal{T}_\omega$ are obtained by applying the Discrete Fourier Transform to $\mathcal{T}_n$. The proportion of the energy concentrated at the driving frequency $\Omega$ (b,c,e,f) as well as at the second most energetic frequency (insets) as a function of $\Omega$ (b,e), the oscillation amplitude $A$ (c,f) for active fluids with the wide (b,c) and small (e,f) active bandwidths $\kappa_\tn{s}$ and $\kappa_\tn{w}$.
}
\label{figSI03} 
\end{figure}

Fig.~\ref{figSI03} quantifies the shape of the power spectral density by displaying the proportion of the PSD concentrated at  $\Omega$ as well as at the second most energetic frequency. Figs.~\ref{figSI03}(a,d) show the full PSD $|\mathcal{T}_\omega|^2$ normalized by the total energy $\sum_\omega |\mathcal{T}_\omega|^2$ for the two simulations shown in Figs.~\ref{fig03}(b,d). Two strong peaks at the driving frequency $\Omega=2\pi/\tau$ confirm that the formula~(\ref{e:linear_response}) holds in these two cases. In general, we measured the proportion of the PSD stored in the driven mode $|\mathcal{T}_\Omega|^2/\sum_\omega |\mathcal{T}_\omega|^2$ [Figs.~\ref{figSI03}(b,c,e,f)] as well as in the second most energetic mode [insets in Figs.~\ref{figSI03}(b,c,e,f)] for different container radii $R$, driving frequencies [Figs.~\ref{figSI03}(b,e)], driving amplitudes [Figs.~\ref{figSI03}(c,f)] and the wide [Figs.~\ref{figSI03}(b,c)] and small [Figs.~\ref{figSI03}(e,f)] active bandwidths $\kappa_\tn{s}$ and $\kappa_\tn{w}$ defined in the Introduction. We found that at least about half of the energy is always concentrated at the driven mode and that this proportion quickly becomes larger than 90\% once the driving frequency $\Omega$ and amplitude $A$ become larger than the corresponding active fluid characteristic pattern formation parameters $2\pi/\tau$ and $U$, respectively. The second most energetic mode typically contains an order or two orders of magnitude less energy than the driven mode. Overall, Fig.~\ref{figSI03} confirms that the response of the torque $\mathcal{T}(t)$ is typically concentrated around the driving frequency $\Omega$, validating the relation~(\ref{e:linear_response}) in the case of active fluids.

\section{Stokes' second problem: dissipative response}
\label{app:stokes_diss_res}
\begin{figure}[t!]
\centering
\includegraphics[width=0.95\textwidth]{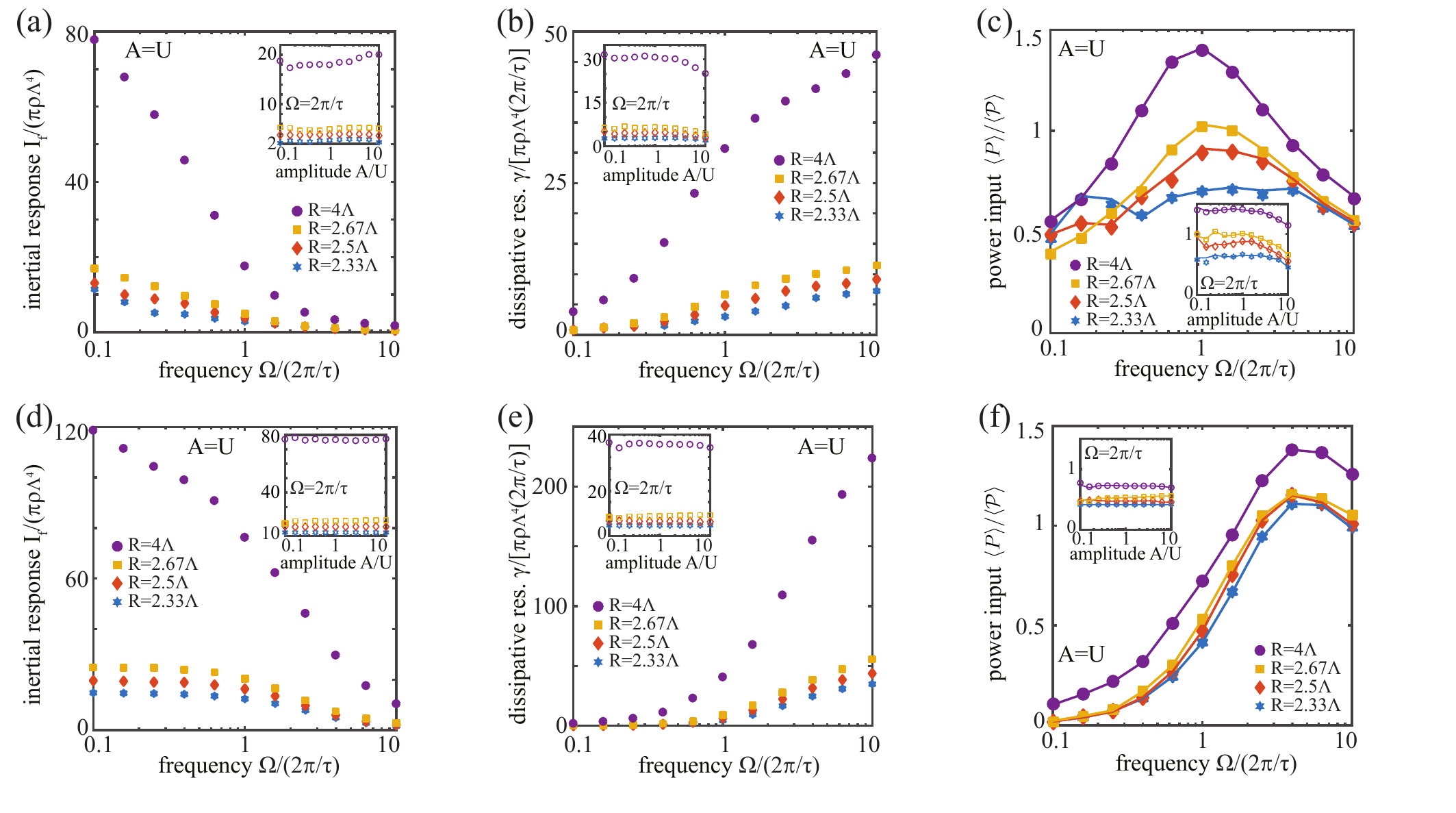}
 \caption{
Inertial (a,d) and dissipative (b,e) response parameters $I_f$ and $\gamma$ as a function of the oscillating frequency $\Omega$ and amplitude $A$ (insets) that appear in the relation~(\ref{e:linear_response}) for active fluids with wide $\kappa_{\tn{w}}=1.5/\Lambda$ (a-c) and small $\kappa_{\tn{s}}=0.63/\Lambda$ (d-f) spectral bandwidths. The parameters were computed using Eq.~(\ref{eq:approx_lin_res}). (c,f) Average power input per unit length $\langle P \rangle$ in the steady state normalized by the value $\langle \mathcal{P} \rangle$ expected from the classical Stokes' problem for a semi-infinite plate shows relative resonance at the characteristic frequency $2\pi/\tau$ of the active flow patterns. The markers indicate power input as computed from the full time series of the torque $\mathcal{T}(t)$ while the lines indicate the contribution derived from the linear relation~(\ref{e:linear_response}).
}
\label{figSI04} 
\end{figure}

The discussion of the active Stokes' second problem in the Main Text focused on the inertial response characterized by the parameter $I_f$ in Eq.~(\ref{e:linear_response}). In this section, we focus on the dissipative response described by the parameter $\gamma$ in that equation. Both, $I_f$ and $\gamma$ are displayed in Fig.~\ref{figSI04}.
\par
Specifically, we are interested in the energy transfer between container and active fluid,  reflected in the average power input per unit length $\langle P \rangle$ needed to sustain the oscillations. As will be shown in the next section, a passive Newtonian fluid ($\Gamma_2=\Gamma_4=0$) such as water confined to a circular container responds effectively as a rigid body under the conditions typical for active fluids experiments. Since an ideal rigid body is a conservative system, we instead benchmark the active fluid dissipative response against the response of a passive Newtonian fluid  filling the upper half-plane and driven horizontally along the $x$-axis. In this classical setting, Stokes' second problem can be solved analytically yielding the power input per unit area  $\langle \mathcal{P}\rangle=\rho A^2 \sqrt{\Omega\Gamma_0/8}$, where $\rho$ and $\Gamma_0$ are the density and kinematic viscosity of the fluid~\cite{landau2013fluid}. Adapting this classical result to thin-films by interpreting $\langle \mathcal{P}\rangle$ as power per unit length and $\rho$ as area density,  $\langle \mathcal{P}\rangle=\rho A^2 \sqrt{\Omega\Gamma_0/8}$ defines a reference for the dissipative response of the active fluid.  
\par
We computed the power input $\langle P \rangle$ in two different ways: using the full time series for the torque $\mathcal{T}(t)$, which gives $\langle P \rangle=\langle \mathcal{T}\dot\phi \rangle$, or approximately, using the relation~(\ref{e:linear_response}), for which $\langle P \rangle\approx \gamma (A/R)^2/2$. To explore how the confinement geometry, driving protocol and active fluid properties affect $\langle P \rangle$ computed in these two ways,  we varied systematically  the amplitude~$A$,  the oscillation frequency~$\Omega$, and the container radius $R$ in our simulations, comparing active fluids with wide  ($\kappa_{\tn{w}}=1.5/\Lambda$) and small ($\kappa_{\tn{s}}=0.63/\Lambda$) spectral bandwidths, respectively. The results of these parameter scans are summarized in Figs.~\ref{figSI04}(c,f). The two ways of computing $\langle P \rangle$, through the exact [markers in Figs.~\ref{figSI04}(c,f)] and approximate [lines in Figs.~\ref{figSI04}(c,f)] formulae, yield almost identical results, further verifying the validity of Eq.~(\ref{e:linear_response}).
\par
Changing the driving amplitude $A$ while keeping the other parameters fixed, we find that the classical power-amplitude scaling $\langle P \rangle\sim A^2$ remains preserved in active fluids to within a good approximation [insets in Figs.~\ref{figSI04}(c,f)].  Our simulations predict,   however,  that passive and active fluids exhibit a fundamentally different response to frequency variations. For both $\kappa\Lambda>1$ and $\kappa\Lambda <1$, we observe deviations from the 1/2 exponent characterized by a relative resonance when the external driving period~$T=2\pi/\Omega$ becomes of the order of the intrinsic vortex growth time scale~$\tau$ [Figs.~\ref{figSI04}(c,f)]. Away from the resonance, the growth is faster than predicted by the $1/2$ exponent at small frequencies and slower than the $1/2$ exponent at large frequencies; the precise growth rates depend on the domain size~$R$, a signature of the interplay between activity and confinement. However, the relative resonance itself is robust against variations in~$R$.

\section{Stokes' second problem: passive fluid response}

\label{app:stokes_passive}
\begin{figure}[h!]
\centering
\includegraphics[width=0.5\textwidth]{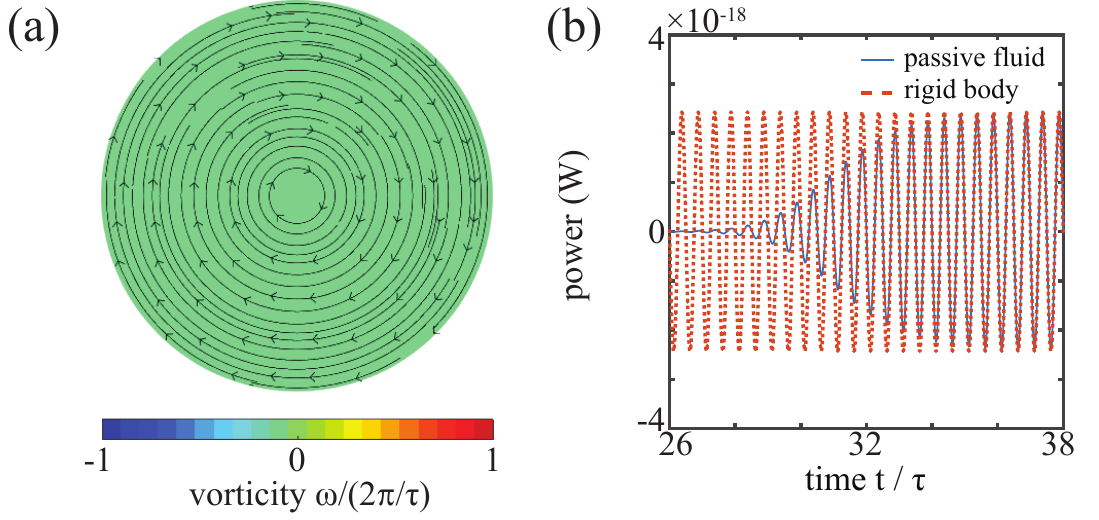}
 \caption{
A passive fluid ($\Gamma_2=\Gamma_4=0$)  with viscosity $\Gamma_0=10^{-6}\tn{m}^2/\tn{s}$ confined to a disk domain of radius $R=200\mu \tn{m}$ subject to oscillatory boundary conditions in Eq.~(\ref{e:stokes_bc}) with angular frequency $\Omega=3.14 \tn{rad}/\tn{s}$ and amplitude $A=628\mu\tn{m}/\tn{s}$ responds effectively as a rigid body. This is because for such parameters, typical for the active Stokes' second problem presented in Fig.~\ref{fig03}, the penetration depth $\delta$ of the passive fluid is much bigger than the domain size $R$. 
(a) Representative snapshot of the vorticity and flow fields illustrates the rigid body-like response.
(b) The corresponding power input (solid line) of the passive fluid driven according to the protocol described in Eq.~(\ref{e:stokes_bc}) follows accurately the formula~(\ref{e:rigid_body_power}) for the power input of a rigid body rotating about the $z$-axis represented by a disk with mass equal to that of the fluid (broken line).
}
\label{figSI05} 
\end{figure}

In this section, we analyze the response of a passive fluid ($\Gamma_2=\Gamma_4=0$) with water viscosity $\Gamma_0=10^{-6}\tn{m}^2/\tn{s}$ to the oscillatory boundary conditions presented in Fig.~\ref{fig03}. We first compare the penetration depth $\delta$~\cite{landau2013fluid}
\be
\delta=\sqrt{2\Gamma_0/\Omega},
\ee
with the domain size $R$. Typical values of the characteristic time scale $\tau$ and vortex size $\Lambda$ at the peak of bacterial activity are $(\tau,\Lambda)=(2\tn{s},50\mu\tn{m})$~\cite{Dunkel2013_PRL}. Therefore, in a potential experiment realizing the set-up in Fig.~\ref{fig03}, one expects frequencies and domain sizes of the order $\Omega\sim2\pi/\tau\sim 3.14\,\tn{rad/s}$, $R\sim 4\Lambda\sim 200\mu\tn{m}$.
For such frequencies, a passive fluid with water viscosity has the penetration depth
\be
\delta\sim 1\tn{mm}\gg R.
\ee
We see that the penetration depth is much bigger than the domain size, which implies that, for the range of domain sizes and driving frequencies relevant to the active Stokes' second problem, the passive fluid effectively responds as a rigid body. Since a rigid body performing harmonic oscillations behaves like a conservative system, one expects a negligible power input in that case. Specifically, a flat disk of radius $R$ and thickness $z$ filled with water with density $\rho=10^{3}\tn{kg}/\tn{m}^3$ has mass $m_\tn{f}=\rho \pi R^2 z$. The corresponding moment of inertia is 
$$I=m_\tn{f}R^2/2=\rho \pi R^4 z/2.$$
The angular speed of the disk is $\dot\phi(t)=(A/R) \cos(\Omega t)$, see Eq.~(\ref{e:stokes_bc}). Then the energy of the rigid disk is $E=I\dot\phi^2/2$. Differentiating with respect to time yields the power of the disk undergoing sinusoidal oscillations about the $z$-axis
\be
\label{e:rigid_body_power}
P=\dot E=I\dot\phi \ddot\phi=-I(A/R)^2\Omega \cos(\Omega t)\sin(\Omega t).
\ee
Averaging the above expression over a period yields zero power input, as expected. We compared this exact expression with the power input for a passive fluid subject to the oscillatory boundary conditions in the disk geometry presented in Fig.~\ref{fig03} with driving parameters $(R,A,\Omega)=(200\mu \tn{m},628\mu\tn{m}/\tn{s},3.14 \tn{rad}/\tn{s})$, typical for the active problem. Fig.~\ref{figSI05} shows the vorticity profile and time series for the power input in a representative simulation. The evolution of the power input is sinusoidal and follows the exact expression~(\ref{e:rigid_body_power}) very closely, implying that the fluid indeed behaves like a rigid body, as expected from the above penetration depth estimates.
\par
The above analysis confirms that a passive fluid responds to the oscillatory boundary conditions in the disk geometry presented in Fig.~\ref{fig03} effectively as a rigid body. In the notation given by Eq.~(\ref{e:linear_response}), the passive response is characterized by $I_\tn{f}=I_\tn{f,p}=m_\tn{f}R^2/2$ and $\gamma_\tn{passive}\approx 0$, justifying the definition of the activity-induced relative added mass $\lambda$ given by $I_\tn{f,a}=(1+\lambda)I_\tn{f,p}$.

\section{Container angular momentum equation}
\label{app:ang_mom_eq}
When a container encapsulating an active fluid is isolated from external forces and torques, it is subject solely to the fluid stresses. The container is taken to be a uniform rigid ring of radius $R$ and mass $m_\tn{c}$. The fluid is assumed to form a planar free standing thin film supported on the ring. Since the fluid is incompressible, the center of mass of the ring is stationary. However, the ring can acquire angular momentum, because the fluid can exert nonzero torque on the container. The ring's angular momentum is 
\be
L_\tn{c}=I\dot{\phi},
\label{eq:ring_ang_mom}
\ee
where $I=m_\tn{c}R^2$ is the moment of inertia and $\dot{\phi}$ is the angular speed. Assuming the ring lies in the $(x,y)$-plane and its center is at the origin, working in polar coordinates $(r,\theta)$, we find that the torque due the fluid stress on a small segment $Rd\theta$ of the ring is
\be
\label{eq:torque_small_element}
\tn{torque on a small segment}=R\hat{\bs r}\times [\rho\bs{\sigma}\cdot(-\hat{\bs r})Rd\theta]=-\rho R^2 \sigma_{r\theta}d\theta,
\ee
where $ \sigma_{r\theta}=\hat{\bs r}\cdot\bs{\sigma}\cdot\hat{\bs \theta}$. The two-dimensional fluid density $\rho$ appears explicitly, since in the Main Text it is our convention that in the stress tensor
\be
\bs\gs=(\Gamma_0 -\Gamma_2 \nabla^2+\Gamma_4 \nabla^4)[\nabla \bs v+ (\nabla \bs v)^\top]
\ee
the parameters $\Gamma_i$ are kinematic quantities. Integrating over the entire ring gives the total torque, and thus the evolution of the ring angular momentum obeys
\be
\f{d}{dt}L_\tn{c}=-\rho R^2\int_0^{2\pi}d\theta \; \sigma_{r\theta}.
\ee
In terms of the ring angular acceleration, we have
\be
m_\tn{c}\ddot{\phi}=-\rho\int_0^{2\pi}d\theta \; \sigma_{r\theta}.
\ee
Non-dimensionalizing as in the first section, we obtain
\be
m_\tn{c}\ddot{\phi}=-\rho R^2\int_0^{2\pi}d\theta\; \sigma_{r\theta},
\ee
where $\bs\gs=(1 -\gamma_2 \nabla^2+\gamma_4 \nabla^4)[\nabla \bs v+ (\nabla \bs v)^\top]$. The fluid mass is $m_\tn{f}=\rho \pi R^2$. We introduce a dimensionless parameter~$\alpha$ representing the ratio of the fluid mass to the ring mass,
\be
\alpha=\f{m_\tn{f}}{m_\tn{c}}.
\ee
We then find that the angular speed of the ring obeys
\be
\ddot{\phi}=-\f{\alpha}{\pi}\int_0^{2\pi}d\theta\; \sigma_{r\theta}.
\ee

\begin{figure}[t!]
\centering
\includegraphics[width=0.95\textwidth]{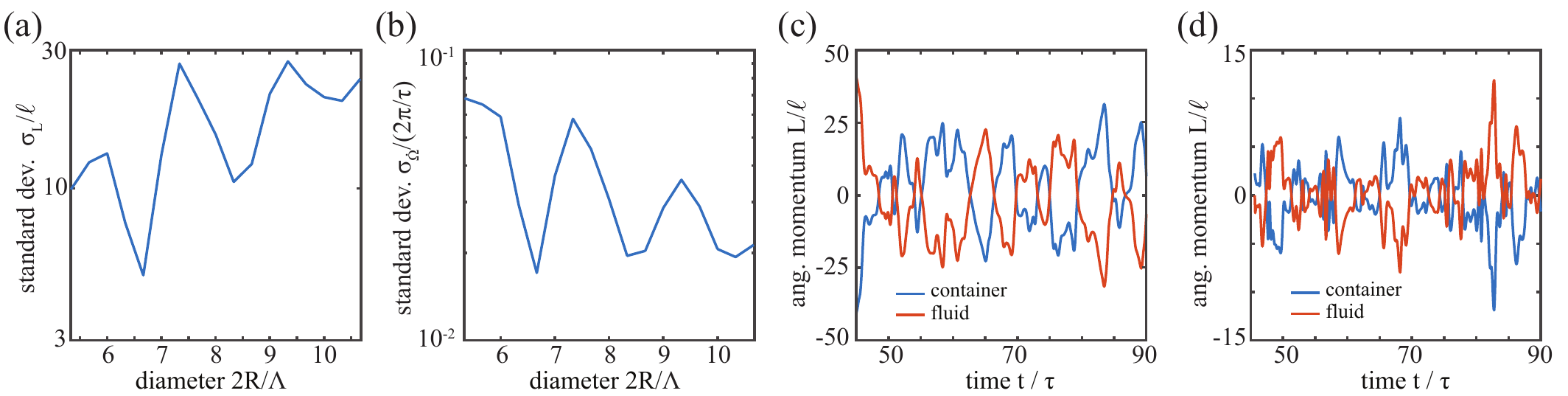}
 \caption{
Geometrically induced oscillatory behavior of fluctuations for an active fluid with narrow vortex-size distribution $\kappa_{\tn{S}}=0.63/\Lambda$. (a) Angular momentum fluctuations $\sigma_L$ as a function of the domain size for heavy containers obtained from Fig.~\ref{fig04}(a) by averaging over $\alpha \in [0.01,0.1]$.
(b) Angular speed fluctuations $\sigma_{\dot{\phi}}$  as a function of the domain size for light containers obtained from Fig.~\ref{fig04}(c) by averaging over $\alpha \geq 10$.
(c, d) Zoom-in of the time series of the container's angular momentum (blue) calculated from Eq.~(\ref{eq:ring_ang_mom}) shown in Fig.~\ref{fig04}(b) for domain radius $R=3.33$ in (c) and $R=3.67$ in (d). 
Additionally, to illustrate the angular momentum conservation in the fluid-container system, we show the  time series of the fluid's angular momentum (orange) calculated independently using the formula $L_{\tn{fluid}}=\rho \int_{0}^R dr\, r\int_0^{2\pi}d\theta \;rv_\theta $.
}
\label{figSI06} 
\end{figure}



%

\end{document}